\documentclass[aps,graphicx,12pt,showkeys,showpacs]{revtex4}
\usepackage{amsmath}
\usepackage{amscd}
\usepackage{graphicx}

\begin{document}

\title[Short Title]{One-step generation of multi-atom Greenberger-Horne-Zeilinger states
in separate cavities via adiabatic passage}
\author{Si-Yang Hao$^{1}$}
\author{Yan Xia$^{1,2,}$\footnote{E-mail: xia-208@163.com}}
\author{Jie Song$^{3}$}
\author{Nguyen Ba An$^{4,}$\footnote{E-mail: nban@iop.vast.ac.vn}}

\affiliation{$^{1}$Department of Physics, Fuzhou University, Fuzhou
50002, China\\$^{2}$School of Physics and Optoelectronic Technology,
Dalian university of Technology, Dalian 116024,
China\\$^{3}$Department of Physics, Harbin Institute of Technology,
Harbin 150001, China\\$^{4}$Center for Theoretical Physics,
Institute of Physics, 10 Dao Tan, Hanoi, Vietnam}

\begin{abstract} We propose a scheme to deterministically generate
Greenberger-Horne-Zeilinger states of $N\geq 3$ atoms trapped in
spatially separated cavities connected by optical fibers. The scheme
is based on the technique of fractional stimulated Raman adiabatic
passage which is one-step in the sense that one needs just wait for
the desired entangled state to be generated in the stationary
regime. The parametrized shapes of the Rabi frequencies of the
classical fields that drive the two end atoms are chosen
appropriately to realize the scheme. We also show numerically that
the proposed scheme is insensitive to the fluctuations of the
pulses' parameters and, at the same time, robust against decoherence
caused by the dissipation due to fiber decay. Moreover, a relatively
high fidelity can be obtained even in the presence of cavity decay
and atomic spontaneous emission.
\end{abstract}


\maketitle

\section{Introduction}

Entanglement is not only an essential ingredient for testing quantum
nonlocality against local hidden theories \cite{JSBP64,DMGMHAS90},
but also a necessary resource for implementing various quantum
informatic tasks. Fundamentally, it is one of the most important
traits in quantum mechanics. It has found different applications in
quantum information processing (QIP) such as quantum cryptography
\cite{AKE91}, quantum teleportation \cite {CHBGBCC93,xiajosab},
quantum dense coding \cite{KMHWPRL96,JWPAZPRA98}, quantum secrete
sharing \cite{MHVBPRA99}, and so on. Typical entangled states are
Bell states \cite{JSBP64}, Greenberger-Horne-Zeilinger (GHZ) states
\cite {DMGMHAS90} and W states \cite{WDGV00} have been identified
and can directly be utilized in QIP
\cite{AKE91,CHBGBCC93,xiajosab,KMHWPRL96,JWPAZPRA98,MHVBPRA99}. In
particular, great interest has been arisen regarding the significant
role of GHZ states in the foundations of quantum mechanics
measurement theory and quantum communication
\cite{SBVVPRA98,RCDGPRL99,VSNG01,GACSDB02,CPYSC04}, error correction
protocols \cite{DPDPRL96,JPPRS98}, and high-precision spectroscopy
\cite{JJBWMI96,SFHCM97}. Contrary to bipartite entangled states, GHZ
states exhibit a special kind of entanglement between $N\geq 3$
parties, providing a possibility to test quantum nonlocality in a
one-shot manner. So, of common interest is the problem of how to
generate GHZ states using current technologies. In fact, for trapped
ions \cite{DLN05} and photons \cite{ZZN04,xiaapl}, a series of
experimental methods \cite {RJNDGCPRA00,DLN05,MNN10} have already
been invented. It is worthy to note that cavity quantum
electrodynamics (CQED) proves to be very promising for QIP. Based on
CQED, numerous schemes \cite
{PBLOE01,XYLPJS09,AZJLAMOP11,SBZEPJD09,XYKLGSPRA09,ASSMPRL06,ZBYSYYAMOP10,ZQYFLLPRA07}
have been proposed for deterministic generation of entanglement
between atoms trapped in different cavities connected by optical
fibers \cite {JICPZPRL97,SVJCPRL97,SBPLK99,SLMSSPRL01,ASPHJKPRA00}.
For example, Zheng \emph{et al.} proposed a simplified scheme to
product GHZ states \cite {ASSJOB99}, while Li \emph{et al.} made use
of the quantum Zeno dynamics \cite{CEPOE12}. Both the schemes are
difficult to implement because they depend on the exact knowledge of
all parameters and require controlling the interaction time
accurately. A way to overcome such difficulties in state engineering
is to force the system's initial state evolve along a dark-state, if
any, by means of adiabatic passage. Such an evolution can be
realized by the so-called technique of stimulated Raman adiabatic
passage
(STIRAP) \cite{PKLTRMP07} or fractional STIRAP (f-STIRAP) \cite{NVVKASJPB99}%
, which have been employed in the context of coherent population transfer
\cite{UGPRJCP90,KBHTRMP98} and coherent atomic beam deflection \cite
{PMPZPRA90}.

In this paper, we design a one-step scheme to deterministically
generate GHZ states of $N$ atoms individually trapped in a linear
array of optical cavities whose nearest neighbors are connected by
$N-1$ optical fibers. Two external lasers are needed to drive the
two end atom-cavity subsystems. The scheme is based on the adiabatic
passage along a dark state, which is a specific eigenstate of the
total atom-cavity-fiber system corresponding to the zero eigenvalue.
The key idea is to choose suitable time-dependent Rabi frequencies
of the driving lasers. Here, we use Gaussian pulses with proper
parameters which are turned on and turned off in an adequate manner
so that in the long-time (stationary) limit the desired GHZ states
emerge automatically. Compared with the previous schemes
\cite{ASSJOB99,CEPOE12}, ours has the following advantages: (i) The
multi-atom GHZ states are produced deterministically only in one
step without worrying about precise control over interaction time;
(ii) The scheme is insensitive to moderate fluctuations of
experimental parameters and (iii) The process is immune to the fiber
decay, and a relatively high fidelity can be obtained even in the
presence of cavity decay and atomic spontaneous emission.

Our paper is organized as follows. After the Introduction, in section II, we
describe the physical model and present the detailed procedure to realize
the scheme for generating three-atom GHZ states. The general case of any $%
N>3 $ atoms is also touched upon briefly in this section. Then, in section
III, we discuss issues regarding the robustness of our scheme against
possible fluctuations in the parameters involved as well as against various
mechanisms of decoherence. Finally, we conclude in section IV.

\begin{figure}[tbp]
\scalebox{0.8}{\includegraphics{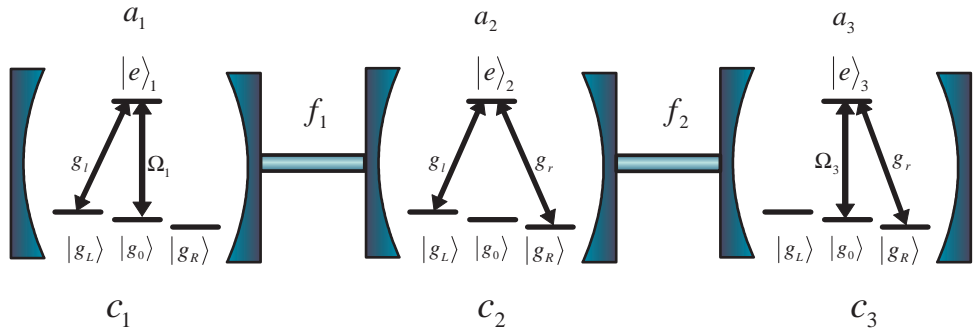}}
\caption{Three atoms ($a_{1}$, $a_{2}$ and $a_{3}$), each of which has one
excited state ($|e\rangle _{k}$ ($k=1,2,3$)) and three ground states ($%
|g_{L}\rangle _{k} $, $|g_{0}\rangle _{k}$ and $|g_{R}\rangle _{k}$) with a
tripod-type configuration, are respectively trapped in three optical
cavities ($c_{1}$, $c_{2}$ and $c_{3}$) connected by two optical fibers ($%
f_{1}$ and $f_{2}$).}
\label{Fig.1}
\end{figure}

\section{Generation of multi-atom GHZ states}

We first consider the case of three atoms in detail. As shown in Fig. 1,
three atoms $a_{1},$ $a_{2}$ and $a_{3}$ are trapped in three distant
linearly arranged optical cavities $c_{1},$ $c_{2}$ and $c_{3},$
respectively. Each atom has one excited state $|e\rangle _{k}$ ($k=1,2,3$)
and three ground states $|g_{L}\rangle _{k},$ $|g_{0}\rangle _{k}$ and $%
|g_{R}\rangle _{k},$ which are nondegenerate and correspond to $J=1$ and $%
m=-1,$ $0,$ $+1,$ respectively. The cavities $c_{1}$ and $c_{3}$ are
single-mode while the cavity $c_{2}$ is two-mode. They are connected by two
short optical fibers $f_{1}$ and $f_{2}.$ The atomic transitions $|e\rangle
_{1(2)}\leftrightarrow |g_{L}\rangle _{1(2)}$ $(|e\rangle
_{2(3)}\leftrightarrow |g_{R}\rangle _{2(3)})$ are resonantly coupled to the
left-circularly (right-circularly) polarized cavity mode, while the
transitions $|e\rangle _{1}\leftrightarrow |g_{0}\rangle _{1}$ $(|e\rangle
_{3}\leftrightarrow |g_{0}\rangle _{3})$ is driven resonantly by an external
classical laser.

\begin{figure}[tbp]
\scalebox{0.8}{\includegraphics{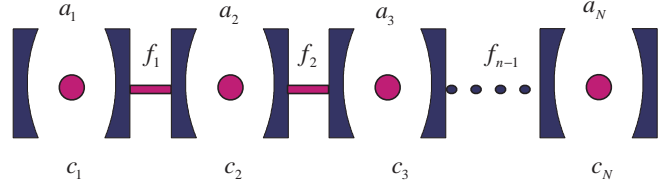}}
\caption{$N$ atoms are respectively trapped in $N$ cavities through $N-1$
fibers.}
\label{Fig.2}
\end{figure}

In the short-fiber limit, $L\nu /(2\pi c)\ll 1$ \cite{TPPRL97,ASSMPRL06}
with $L$ the fiber length, $c$ the speed of light and $\nu $ the decay rate
of the cavity field into a continuum of fiber modes, only one resonant fiber
mode interacts with the cavity mode. Then, in the interaction picture, the
Hamiltonian of the total atom-cavity-fiber system can be written as $(\hbar
=1)$%
\begin{equation}
H=H_{al}+H_{ac}+H_{cf},  \label{H}
\end{equation}
\begin{equation}
H_{al}=\Omega _{1}(t)e^{i\varphi _{1}}|e\rangle _{1}\langle g_{0}|+\Omega
_{3}(t)e^{i\varphi _{3}}|e\rangle _{3}\langle g_{0}|+H.c.,  \label{hl}
\end{equation}
\begin{equation}
H_{ac}=\sum_{i=1}^{2}g_{i,l}a_{i,l}|e\rangle _{i}\langle
g_{L}|+\sum_{i=2}^{3}g_{i,r}a_{i,r}|e\rangle _{i}\langle g_{R}|+H.c.,
\label{hc}
\end{equation}
\begin{equation}
H_{cf}=v_{1}b_{1}^{\dag }(a_{1,l}+a_{2,l})+v_{2}b_{2}^{\dag
}(a_{2,r}+a_{3,r})+H.c.,  \label{hcf}
\end{equation}
with $\Omega _{1(3)}(t)$ and $\varphi _{1(3)}$ the Rabi frequencies and
phases of the driving lasers, $a_{i,l(r)}^{\dagger }$ and $a_{i,l(r)}$ the
creation and annihilation operators for the left-circularly
(right-circularly) polarized mode of cavity $c_{i},$ and $b_{f}^{\dag }$ $%
(b_{f})$ the creation (annihilation) operators of the resonant mode of fiber
$f_{f}.$ For simplicity, we assume equal atom-cavity and equal cavity-fiber
coupling strengths, i.e., $g_{i,l}=g_{i,r}=g$ and $v_{1}=v_{2}=v.$ We also
denote by $|n\rangle _{c1(c3)}$ the quantum field state of cavity $c_{1}$ $%
(c_{3})$ containing $n$ left-circularly (right-circularly) polarized
photons, by $|m,n\rangle _{c2}$ the quantum field state of cavity $c_{2}$
containing $m$ left-circularly and $n$ right-circularly polarized photons,
and by $|n\rangle _{f1(f2)}$ the quantum field state of fiber $f_{1}$ $%
(f_{2})$ containing $n$ photons.

Let the total system be initially in the separable state
\begin{equation}
|\Psi (-\infty )\rangle =|g_{0},g_{L},g_{R}\rangle
_{a_{1}a_{2}a_{3}}|0\rangle _{c1}|0\rangle _{f1}|0,0\rangle _{c2}|0\rangle
_{f2}|0\rangle _{c3}.  \label{Psitvc}
\end{equation}
Then, governed by the Hamiltonian in Eq. (\ref{H}), it will evolve in a
closed subspace spanned by $11$ basis states $\{|\phi _{l}\rangle
;l=1,2,...,11\}:$%
\begin{equation}
|\phi _{1}\rangle =|\Psi (-\infty )\rangle
=|g_{0},g_{L},g_{R}\rangle _{a_{1}a_{2}a_{3}}|0\rangle
_{c1}|0\rangle _{f1}|0,0\rangle _{c2}|0\rangle _{f2}|0\rangle _{c3},
\label{p1}
\end{equation}
\begin{equation}
|\phi _{2}\rangle =|e,g_{L},g_{R}\rangle _{a_{1}a_{2}a_{3}}|0\rangle
_{c1}|0\rangle _{f1}|0,0\rangle _{c2}|0\rangle _{f2}|0\rangle _{c3},
\end{equation}
\begin{equation}
|\phi _{3}\rangle =|g_{L},g_{L},g_{R}\rangle _{a_{1}a_{2}a_{3}}|1\rangle
_{c1}|0\rangle _{f1}|0,0\rangle _{c2}|0\rangle _{f2}|0\rangle _{c3},
\end{equation}
\begin{equation}
|\phi _{4}\rangle =|g_{L},g_{L},g_{R}\rangle _{a_{1}a_{2}a_{3}}|0\rangle
_{c1}|1\rangle _{f1}|0,0\rangle _{c2}|0\rangle _{f2}|0\rangle _{c3},
\end{equation}
\begin{equation}
|\phi _{5}\rangle =|g_{L},g_{L},g_{R}\rangle _{a_{1}a_{2}a_{3}}|0\rangle
_{c1}|0\rangle _{f1}|1,0\rangle _{c2}|0\rangle _{f2}|0\rangle _{c3},
\end{equation}
\begin{equation}
|\phi _{6}\rangle =|g_{L},e,g_{R}\rangle _{a_{1}a_{2}a_{3}}|0\rangle
_{c1}|0\rangle _{f1}|0,0\rangle _{c2}|0\rangle _{f2}|0\rangle _{c3},
\end{equation}
\begin{equation}
|\phi _{7}\rangle =|g_{L},g_{R},g_{R}\rangle _{a_{1}a_{2}a_{3}}|0\rangle
_{c1}|0\rangle _{f1}|0,1\rangle _{c2}|0\rangle _{f2}|0\rangle _{c3},
\end{equation}
\begin{equation}
|\phi _{8}\rangle =|g_{L},g_{R},g_{R}\rangle _{a_{1}a_{2}a_{3}}|0\rangle
_{c1}|0\rangle _{f1}|0,0\rangle _{c2}|1\rangle _{f2}|0\rangle _{c3},
\end{equation}
\begin{equation}
|\phi _{9}\rangle =|g_{L},g_{R},g_{R}\rangle _{a_{1}a_{2}a_{3}}\left|
0\right\rangle _{c1}|0\rangle _{f1}|0,0\rangle _{c2}|0\rangle _{f2}|1\rangle
_{c3},
\end{equation}
\begin{equation}
|\phi _{10}\rangle =|g_{L},g_{R},e\rangle _{a_{1}a_{2}a_{3}}\left|
0\right\rangle _{c1}|0\rangle _{f1}|0,0\rangle _{c2}|0\rangle _{f2}|0\rangle
_{c3},
\end{equation}
and
\begin{equation}
|\phi _{11}\rangle =|g_{L},g_{R},g_{0}\rangle _{a_{1}a_{2}a_{3}}|0\rangle
_{c1}|0\rangle _{f1}|0,0\rangle _{c2}|0\rangle _{f2}|0\rangle _{c3}.
\label{p11}
\end{equation}
The Hamiltonian (\ref{H}), in terms of the basis states Eqs. (\ref{p1} - \ref
{p11}), reads
\begin{equation}
{H}=\left( {\
\begin{array}{ccccccccccc}
0 & \widetilde{\Omega }_{1}(t) & 0 & 0 & 0 & 0 & 0 & 0 & 0 & 0 & 0 \\
\widetilde{\Omega }_{1}^{*}(t) & 0 & g & 0 & 0 & 0 & 0 & 0 & 0 & 0 & 0 \\
0 & g & 0 & v & 0 & 0 & 0 & 0 & 0 & 0 & 0 \\
0 & 0 & v & 0 & v & 0 & 0 & 0 & 0 & 0 & 0 \\
0 & 0 & 0 & v & 0 & g & 0 & 0 & 0 & 0 & 0 \\
0 & 0 & 0 & 0 & g & 0 & g & 0 & 0 & 0 & 0 \\
0 & 0 & 0 & 0 & 0 & g & 0 & v & 0 & 0 & 0 \\
0 & 0 & 0 & 0 & 0 & 0 & v & 0 & v & 0 & 0 \\
0 & 0 & 0 & 0 & 0 & 0 & 0 & v & 0 & g & 0 \\
0 & 0 & 0 & 0 & 0 & 0 & 0 & 0 & g & 0 & \widetilde{\Omega }_{3}(t) \\
0 & 0 & 0 & 0 & 0 & 0 & 0 & 0 & 0 & \widetilde{\Omega }_{3}^{*}(t) & 0
\end{array}
}\right) ,  \label{h}
\end{equation}
with $\widetilde{\Omega }_{1(3)}(t)\equiv \Omega _{1(3)}(t)e^{-i\varphi
_{1(3)}}.$

An alternative basis of the subspace consists of $11$ time-dependent
eigenstates of the instantaneous Hamiltonian given by Eq. (\ref{h}) $%
\{\left| \Phi _{m}(t)\right\rangle ,m=1,2,...,11\}.$ It can be verified that
one of the eigenvalues of such $H$ is zero and the corresponding eigenstate,
which we label by $\left| \Phi _{1}(t)\right\rangle ,$ has the form
\begin{eqnarray}
\left| \Phi _{1}(t)\right\rangle &=&\frac{G(t)}{\sqrt{%
4X^{2}(t)+G^{2}(t)[X^{2}(t)+1]}}|\phi _{1}\rangle  \nonumber \\
&&-\frac{X(t)}{\sqrt{4X^{2}(t)+G^{2}(t)[X^{2}(t)+1]}}\left( |\phi
_{3}\rangle -|\phi _{5}\rangle +|\phi _{7}\rangle -|\phi _{9}\rangle \right)
\nonumber \\
&&-\frac{e^{i(\varphi _{1}+\varphi _{3})}G(t)X(t)}{\sqrt{%
4X^{2}(t)+G^{2}(t)[X^{2}(t)+1]}}|\phi _{11}\rangle ,  \label{Phi1}
\end{eqnarray}
where $X(t)=\Omega _{1}(t)/\Omega _{3}(t)$ and $G(t)=g/\Omega _{3}(t).$ The
state $\left| \Phi _{1}(t)\right\rangle $ is called a trapped or dark state
since it contains atoms only in the ground states during the entire
evolution. The fact that excited levels $\left| e\right\rangle _{k}$ are
missing in $\left| \Phi _{1}(t)\right\rangle $ is due to the destructive
quantum interference. Namely, as seen from Eqs. (\ref{p1} - \ref{p11}), $%
\left| \phi _{2}\right\rangle $ $(\left| \phi _{6}\right\rangle ,$ $\left|
\phi _{10}\right\rangle )$ does contain the excited level $\left|
e\right\rangle _{1}$ $(\left| e\right\rangle _{2},$ $\left| e\right\rangle
_{3}),$ but, the transition $\left| \phi _{1}\right\rangle \rightarrow
\left| \phi _{2}\right\rangle $ $(\left| \phi _{5}\right\rangle \rightarrow
\left| \phi _{6}\right\rangle ,$ $\left| \phi _{9}\right\rangle \rightarrow
\left| \phi _{10}\right\rangle )$ is canceled by the transition $\left| \phi
_{3}\right\rangle \rightarrow \left| \phi _{2}\right\rangle $ $(\left| \phi
_{7}\right\rangle \rightarrow \left| \phi _{6}\right\rangle ,$ $\left| \phi
_{11}\right\rangle \rightarrow \left| \phi _{10}\right\rangle ).$ It is also
of a surprise that the fibers' modes $b_{1}$ and $b_{2}$ do not appear in $%
\left| \Phi _{1}(t)\right\rangle $ (i.e., $\left| \phi _{4}\right\rangle $
and $\left| \phi _{8}\right\rangle $ that contain a fiber mode are absent in
$\left| \Phi _{1}(t)\right\rangle ).$ This is again a consequence of
destructive quantum interference: the transitions $\left| \phi
_{3}\right\rangle \rightarrow \left| \phi _{4}\right\rangle $ $(\left| \phi
_{7}\right\rangle \rightarrow \left| \phi _{8}\right\rangle )$ and $\left|
\phi _{5}\right\rangle \rightarrow \left| \phi _{4}\right\rangle $ $(\left|
\phi _{9}\right\rangle \rightarrow \left| \phi _{8}\right\rangle )$ destroy
each other completely. However, both the atomic excited levels and the
fibers' modes are important, via their virtual excitations, in connecting
the different atomic ground levels as well as the different cavities, a very
necessary feature in our system to generate entanglement between separated
atoms.

The total atom-cavity-fiber system state $\left| \Psi (t)\right\rangle $ at
any time $t$ can be extended as a superposition of $\{\left| \Phi
_{m}(t)\right\rangle \},$%
\begin{equation}
\left| \Psi (t)\right\rangle =\sum_{m=1}^{11}w_{m}(t)\left| \Phi
_{m}(t)\right\rangle ,\text{ }\sum_{m=1}^{11}|w_{m}(t)|^{2}=1,  \label{Psit}
\end{equation}
with $|w_{m}(t)|^{2}$ the probability of finding the system in state $\left|
\Phi _{m}(t)\right\rangle .$ If the pulse that drives the atom $a_{3}$
precedes the pulse that drives the atom $a_{1},$ i.e.,
\begin{equation}
\lim_{t\rightarrow -\infty }\Omega _{3}(t)>\lim_{t\rightarrow -\infty
}\Omega _{1}(t)=0,  \label{con1}
\end{equation}
or $\lim_{t\rightarrow -\infty }X(t)=X(-\infty )=0,$ then from Eqs. (\ref
{Phi1}), (\ref{Psit}) and (\ref{con1}) it follows that $\left| \Phi
_{1}(-\infty )\right\rangle =\left| \phi _{1}\right\rangle =\left| \Psi
(-\infty )\right\rangle .$ Using this in Eq. (\ref{Psit}) yields $%
w_{m=1}(-\infty )=1$ and $w_{m>1}(-\infty )=0.$ If we vary the
pulses slowly enough in time to satisfy the adiabatic following
condition, then the system initial state should evolve only along
$\left| \Phi _{1}(t)\right\rangle $ of Eq. (\ref{Phi1}). The key
strategy is to tailor the pulses so that
\begin{equation}
\frac{2\Omega _{1}(t)\Omega _{3}(t)}{\sqrt{\Omega _{1}^{2}(t)+\Omega
_{3}^{2}(t)}}\ll g,  \label{g}
\end{equation}
(or $2X(t)/\sqrt{X^{2}(t)+1}\ll G(t))$ to keep the probabilities of finding
the system in states $\{|\phi _{3}\rangle ,|\phi _{5}\rangle ,|\phi
_{7}\rangle ,|\phi _{9}\rangle \}$ negligible during the evolution and in
the long-time limit both the pulses vanish simultaneously retaining their
ratio constant, i.e.,
\begin{equation}
\lim_{t\rightarrow \infty }\Omega _{1,3}(t)=0;\text{ }\lim_{t\rightarrow
\infty }[\Omega _{1}(t)/\Omega _{3}(t)]=\lim_{t\rightarrow \infty
}X(t)=X(\infty )=const.  \label{con2}
\end{equation}
If so
\begin{equation}
\lim_{t\rightarrow \infty }\left| \Phi _{1}(t)\right\rangle =\left|
ghz\right\rangle _{123}\otimes |0\rangle _{c1}|0\rangle _{f1}|0,0\rangle
_{c2}|0\rangle _{f2}|0\rangle _{c3},
\end{equation}
with
\begin{equation}
\left| ghz\right\rangle _{123}=\frac{1}{\sqrt{X^{2}(\infty )+1}}%
|g_{0},g_{L},g_{R}\rangle _{a_{1}a_{2}a_{3}}-e^{i(\varphi _{1}+\varphi _{3})}%
\frac{X(\infty )}{\sqrt{X^{2}(\infty )+1}}|g_{L},g_{R},g_{0}\rangle
_{a_{1}a_{2}a_{3}},
\end{equation}
which for $X(\infty )=1$ and $\varphi _{1}+\varphi _{3}=\pi $ is the desired
GHZ state $\left| GHZ\right\rangle _{123}=(|g_{0},g_{L},g_{R}\rangle
+|g_{L},g_{R},g_{0}\rangle )_{a_{1}a_{2}a_{3}}/\sqrt{2}.$

We next briefly present the generalization of the above scheme to the case
of $N>3$ atoms. As Fig. 2 shows, the $N$ atoms $a_{1},a_{2},$ ..., $a_{N}$
are respectively trapped in $N$ cavities $c_{1},c_{2},$ ..., $c_{N}$
connected by $N-1$ fibers $f_{1},f_{2},$ ..., $f_{N-1}.$ The level
configurations of atoms $a_{1},$ $\{a_{2},a_{3},...,a_{N-1}\}$ and $a_{N}$
are the same as those of atoms $a_{1},$ $a_{2}$ and $a_{3}$ in the case of $%
N=3.$ For example, for an odd $N>3$ and equal atom-cavity and equal
cavity-fiber coupling strengths, the expressions of $H_{al},$ $H_{ac}$ and $%
H_{cf}$ in the total Hamiltonian $H$ read

\begin{equation}
H_{al}=\Omega _{1}(t)e^{i\varphi _{1}}|e\rangle _{1}\langle g_{0}|+\Omega
_{N}(t)e^{i\varphi _{N}}|e\rangle _{N}\langle g_{0}|+H.c.,  \label{hlN}
\end{equation}
\begin{equation}
H_{ac}=g\left[ \sum_{i=1}^{N-1}a_{i,l}|e\rangle _{i}\langle
g_{L}|+\sum_{i=2}^{N}a_{i,r}|e\rangle _{i}\langle g_{R}|\right] +H.c.,
\label{hcN}
\end{equation}
\begin{equation}
H_{cf}=v\sum_{i=1}^{(N-1)/2}[b_{2i-1}^{\dag
}(a_{2i-1,l}+a_{2i,l})+b_{2i}^{\dag }(a_{2i,r}+a_{2i+1,r})]+H.c..
\label{hcfN}
\end{equation}
Suppose that initially the atoms are prepared in the separable state $%
|g_{0},g_{L},g_{R},g_{L},\ldots ,g_{R}\rangle _{a_{1}a_{2}\ldots a_{N}}$
while all the cavities and fibers are empty. Then, under the constraint $%
2\Omega _{1}(t)\Omega _{N}(t)/\sqrt{\Omega _{1}^{2}(t)+\Omega _{N}^{2}(t)}%
\ll g$ and the adiabatic following condition, the atoms can eventually
appear in the entangled state $|ghz\rangle _{a_{1}a_{2}\ldots a_{N}}=\cos
\alpha |g_{0},g_{L},g_{R},g_{L},\ldots ,g_{R}\rangle +e^{i(\varphi
_{1}+\varphi _{N})}\sin \alpha |g_{L},g_{R},g_{L},g_{R},\ldots ,g_{0}\rangle
)_{a_{1}a_{2}\ldots a_{N}},$ if $\lim_{t\rightarrow -\infty }\Omega
_{N}(t)>\lim_{t\rightarrow -\infty }\Omega _{1}(t)=0,$ $\lim_{t\rightarrow
\infty }\Omega _{1,N}(t)=0$ and $\lim_{t\rightarrow \infty }[\Omega
_{1}(t)/\Omega _{N}(t)]=\tan \alpha .$

\section{Realization and discussion}

A possible implementation of our scheme for $N=3$ can be realized by using
driving pulses with the Rabi frequencies of the shapes \cite{NVVKASJPB99}
\begin{equation}
\Omega _{1}(t)=\Omega _{0}\sin \alpha \text{ }e^{-(t-\tau )^{2}/T^{2}},
\label{o1}
\end{equation}
and
\begin{equation}
\Omega _{3}(t)=\Omega _{0}e^{-(t+\tau )^{2}/T^{2}}+\Omega _{0}\cos \alpha
\text{ }e^{-(t-\tau )^{2}/T^{2}}.  \label{o3}
\end{equation}
For illustration, we display in Fig. 3 the time-dependence of the pulses $%
\Omega _{1,3}(t),$ the probabilities $P_{n}(t)=|\left\langle \phi
_{n}\right. \left| \Phi _{1}(t)\right\rangle |^{2}$ of finding the system in
the states $\left| \phi _{n}\right\rangle $ $(n=1,3,...,11)$ and the
fidelity $F=|\left\langle GHZ\right. \left| \Phi _{1}(t)\right\rangle |^{2}$
with the parameters $\alpha =\pi /4,$ $\Omega _{0}=0.1g,$ $\tau =50/g$ and $%
T=80/g.$ As is clear from the top figure, the pulses (\ref{o1}) and (\ref{o3}%
) with the above-chosen parameters satisfy the conditions Eq. (\ref{con1})
and Eq. (\ref{con2}) with $X(\infty )=\tan \alpha .$ The middle figure shows
that $P_{1}$ is decreasing from $1$ to $0.5,$ $P_{11}$ is instead increasing
from $0$ to $0.5,$ while $P_{3}=P_{5}=P_{7}=P_{9}$ remains negligible all
the time (in fact $P_{3,5,7,9}$ is increasing from $0$ until a maximum value
of about $0.0032$ when $gt\simeq 38$ but then is decreasing back to $0).$ In
theory, the GHZ state is generated asymptotically in the long-time limit.
However, as seen from the bottom figure, at $gt=100$ the fidelity is already
$F(gt=100)\simeq 0.99$ and at $gt=170$ it is almost unity: $F(gt=170)\simeq
0.9999.$

\begin{figure}[tbp]
\scalebox{0.8}{\includegraphics{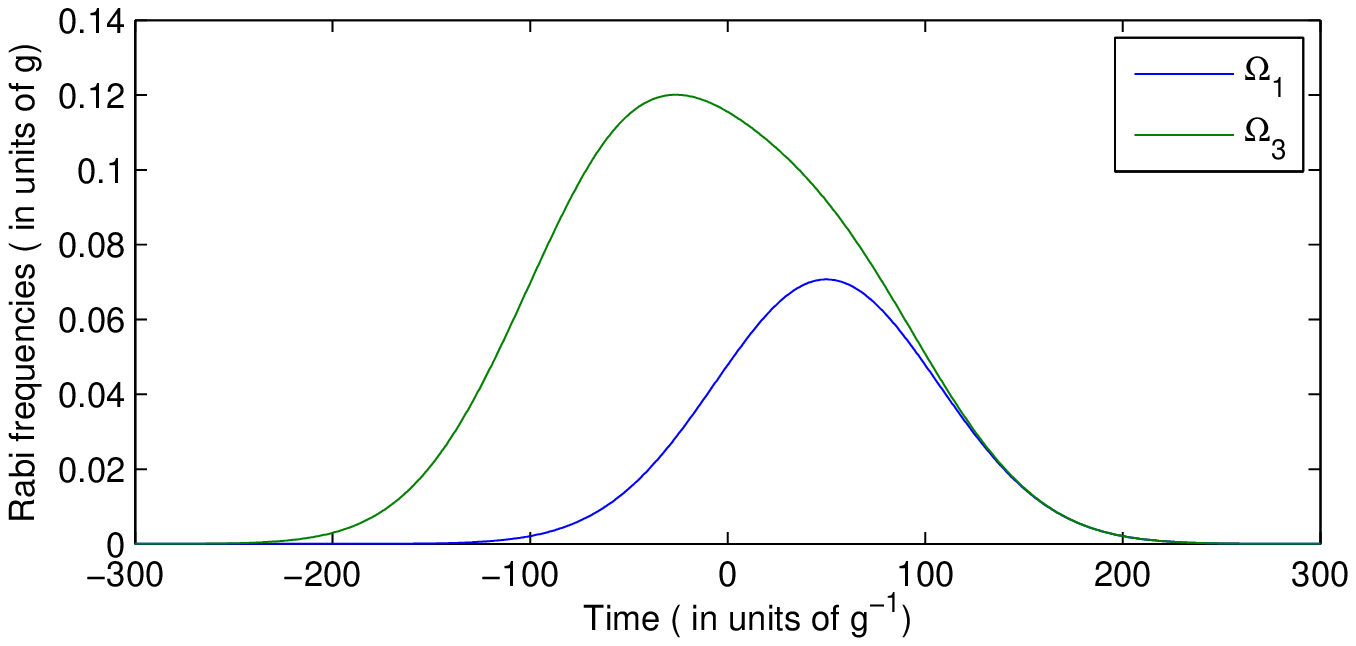}}
\par
\scalebox{0.8}{\includegraphics{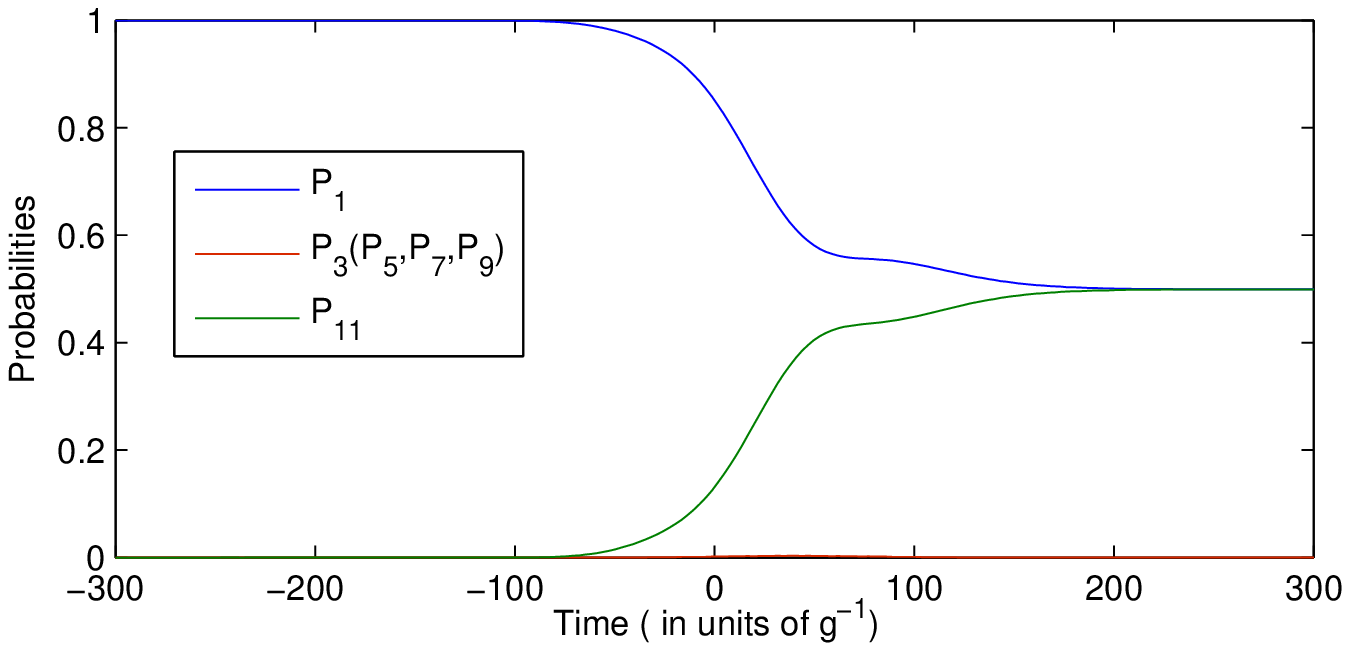}}
\par
\scalebox{0.8}{\includegraphics{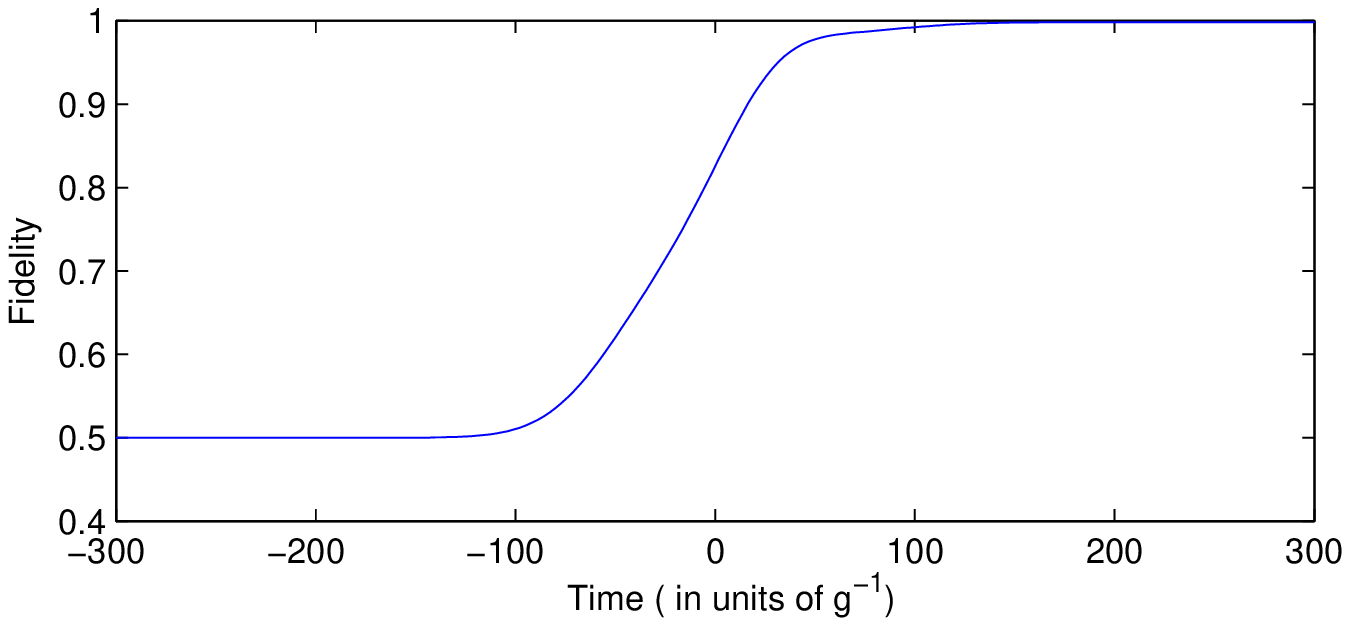}}
\caption{The time dependence of (top) the Rabi frequencies of the driving
lasers $\Omega_1(t)$ and $\Omega_3(t),$ (middle) the probabilities $P_1,
P_3, P_5, P_7, P_9, P_{11}$ of finding the system in states $|\phi_1\rangle$%
, $|\phi_3\rangle$, $|\phi_5\rangle$, $|\phi_7\rangle$, $|\phi_9\rangle$, $%
|\phi_{11}\rangle$, respectively, and (bottom) the fidelity $F.$ The
parameters used are $\alpha =\pi /4,$ $\Omega _{0}/g=0.1,$ $g\tau =50$ and $%
gT=80.$}
\label{Fig.3}
\end{figure}

\begin{figure}[tbp]
\scalebox{0.8}{\includegraphics{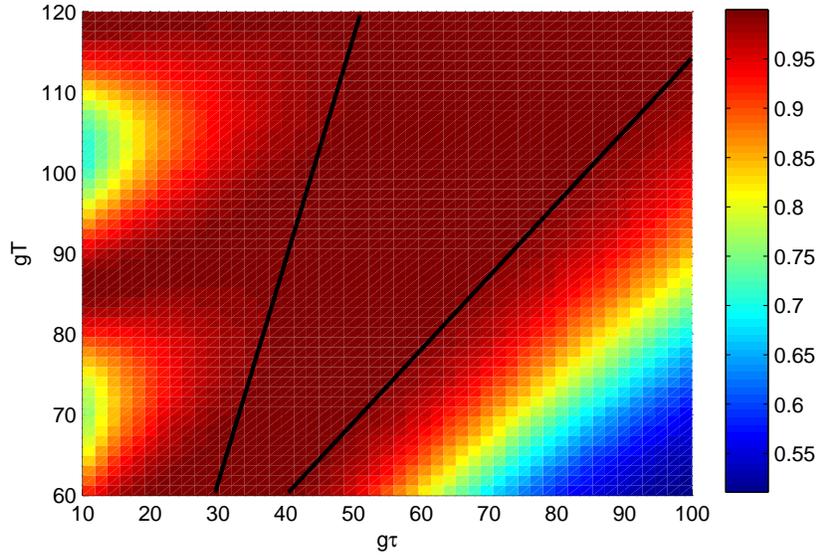}}
\caption{Density plot of the fidelity $F$ at $gt=300$ as a function of $g\tau
$ and $gT$ for $\Omega_0/g=0.1$ and $v/g=10.$ The two straight lines
indicate the boundaries of the region within which the fidelity is almost
one.}
\label{Fig.4}
\end{figure}

\begin{figure}[tbp]
\scalebox{0.8} {\includegraphics{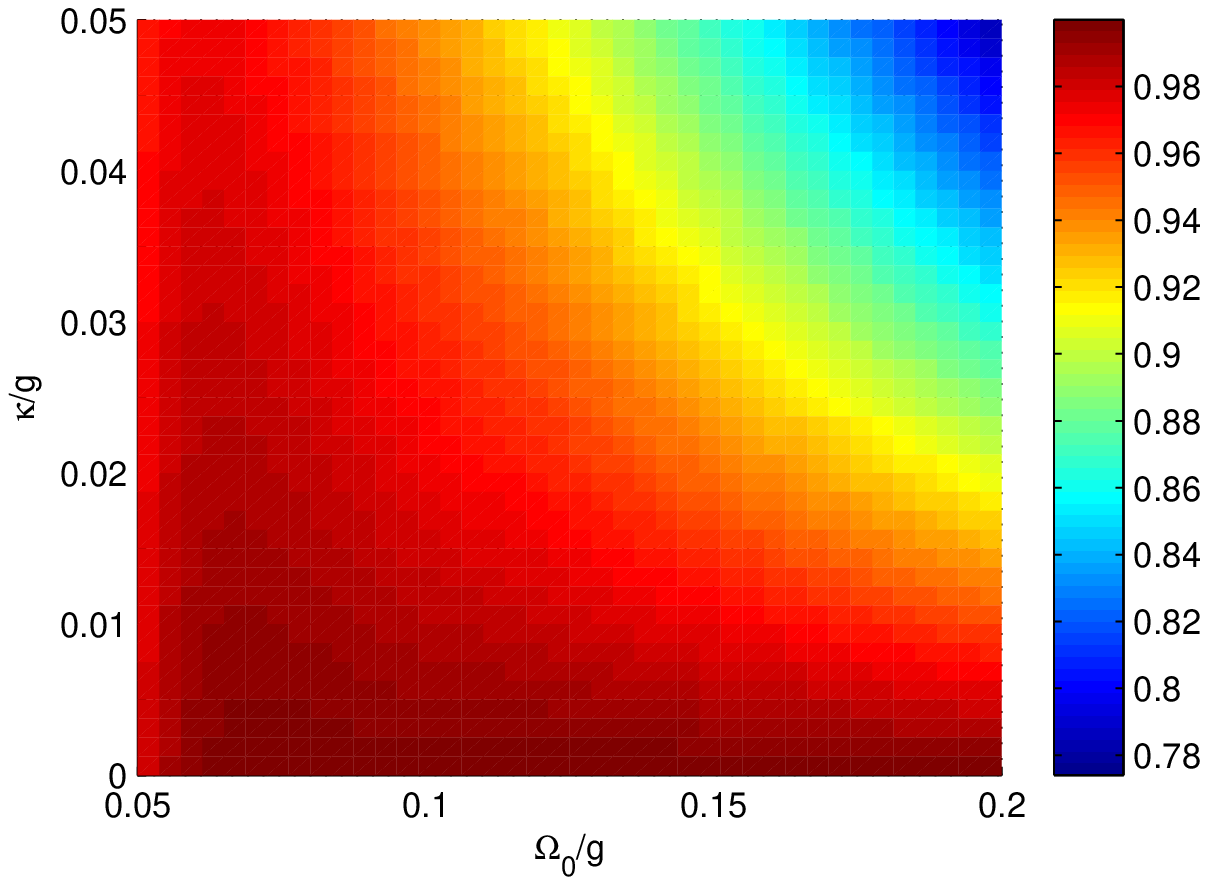}}
\par
\scalebox{0.8} {\includegraphics{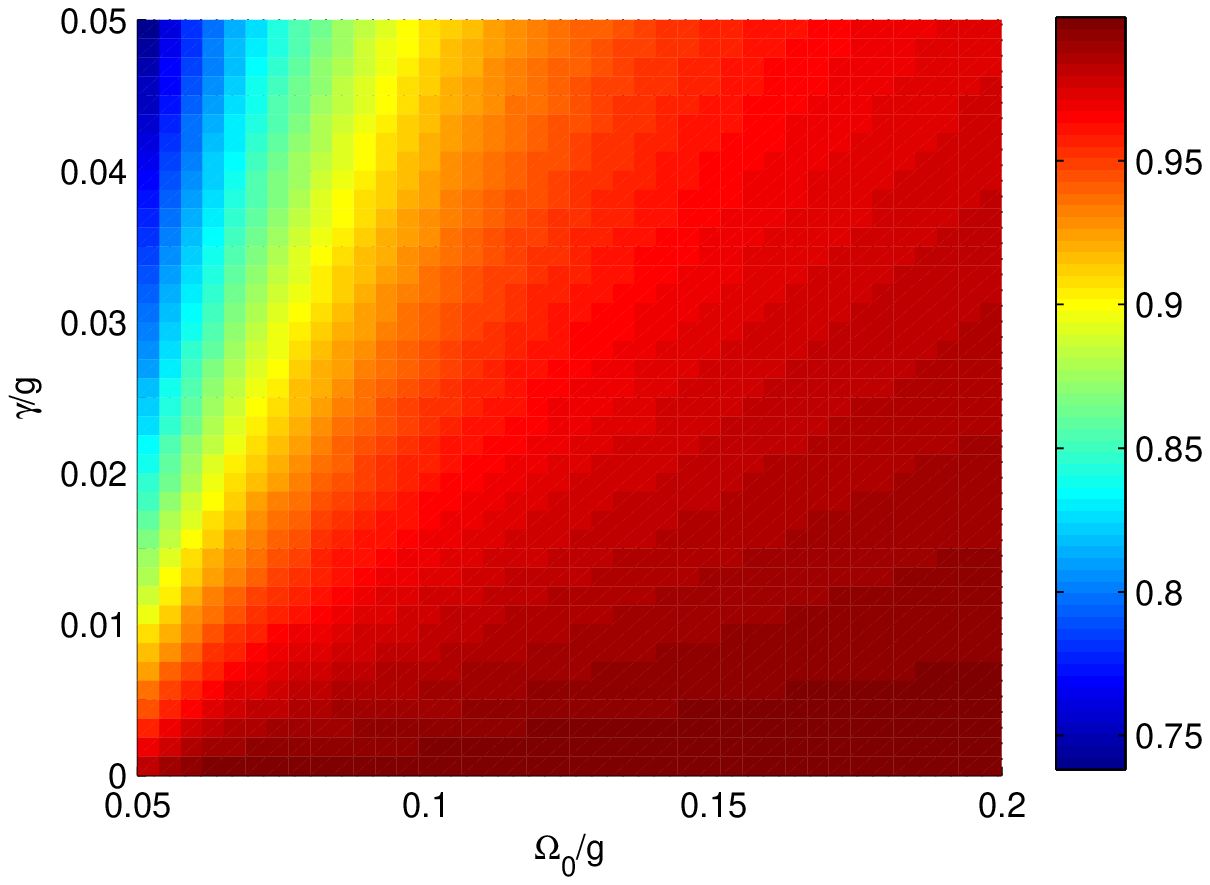}}
\caption{Density plot of the fidelity $F$ at $gt=300$ as a function of (top)
$\Omega_0/g$ and $\kappa/g$ and (bottom) $\Omega_0/g$ and $\gamma/g.$}
\label{Fig.5}
\end{figure}

\begin{figure}[tbp]
\scalebox{0.8}{\includegraphics{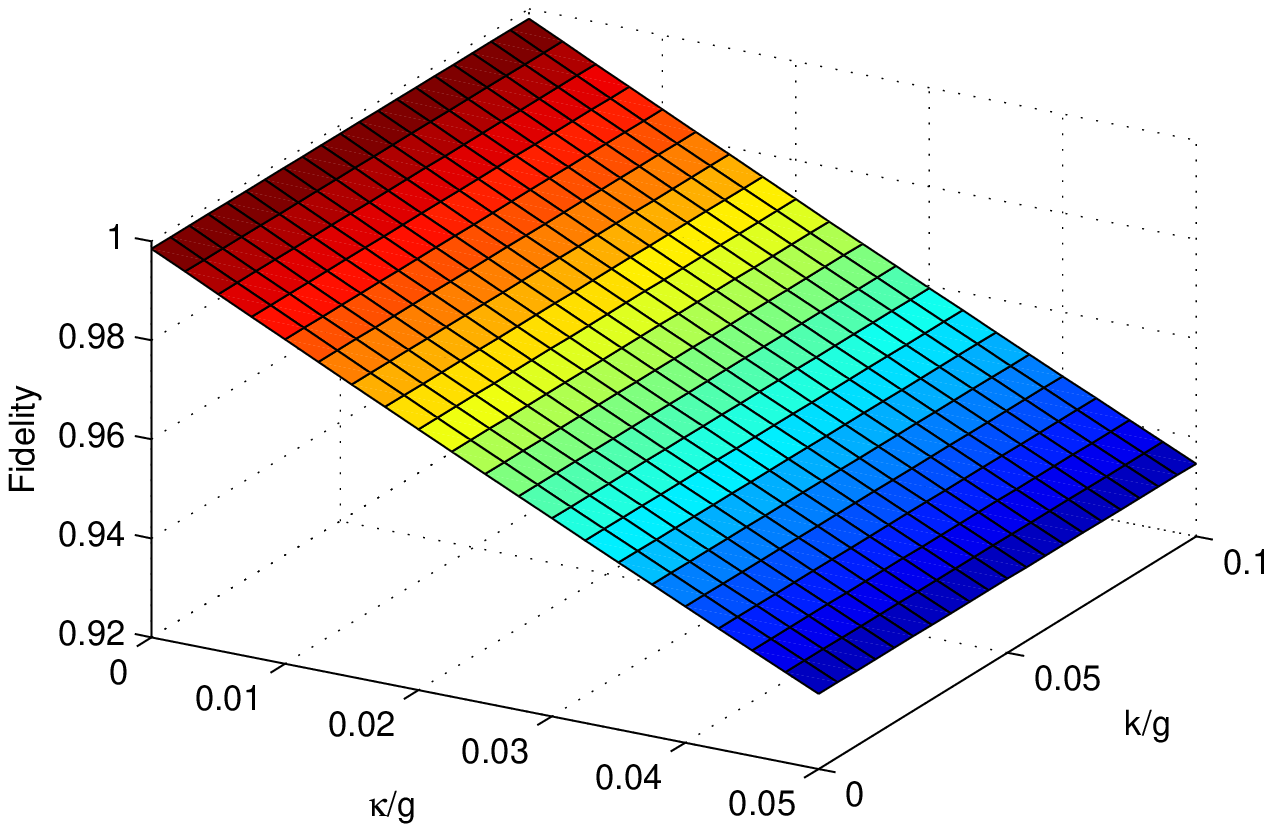}}
\caption{The fidelity $F$ as a function of $\kappa/g$ and $k/g$ for $%
\Omega_0/g=0.1$, $v/g=10$ and $gt=300$.}
\label{Fig.6}
\end{figure}

\begin{figure}[tbp]
\scalebox{0.8}{\includegraphics{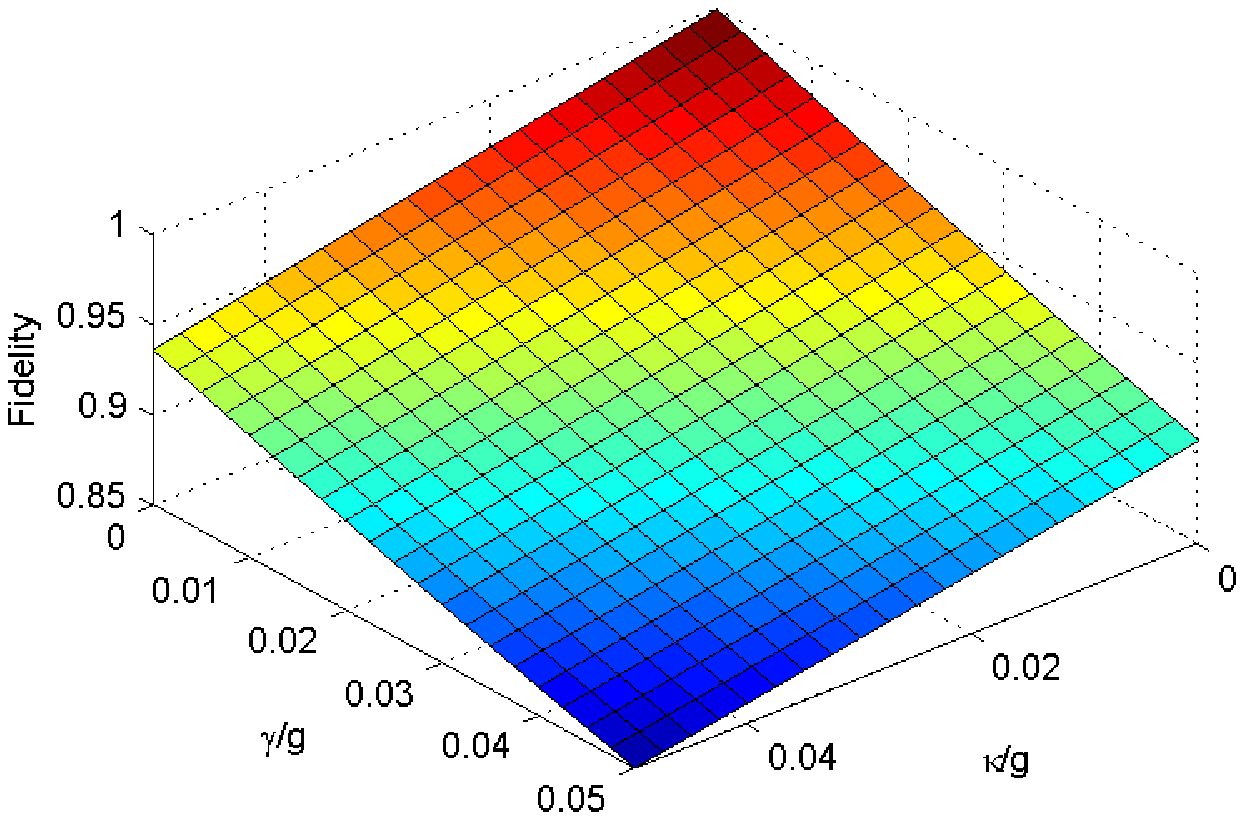}}
\caption{The fidelity $F$ as a function of $\kappa/g$ and $\gamma/g$ for $%
\Omega_0/g=0.1$, $v/g=10$ and $gt=300$.}
\label{Fig.7}
\end{figure}

Next, we will analyze the robustness of adiabaticity condition against the
pulse shapes of classical fields. From Ref. \cite{NVVKASJPB99}, we know that
our scheme needs an optimal range of $\tau $ related to $T$ to achieve a
preferable adiabaticity. So during the evolution, a major challenge is to
choose an optimal relation between $\tau $ and $T.$ The dependence of the
fidelity $F$ on the pulses' parameters shown in Fig. 4 indicates that $F$ is
larger than $99\%$ for $\tau /T$ within the range $0.37\leq \tau /T\leq
1.11. $ This means that our scheme could content with the preferable
adiabaticity in a relatively large range.

So far the whole system is treated as absolutely isolated from the
environment, i.e., we have totally omitted the decoherence effect in our
system. In order to confirm the validity of our scheme, we now discuss on
the influence of decoherence induced by cavity decay, fiber decay and atomic
spontaneous emission. To account for the decoherence we resort to the master
equation for the density matrix $\rho (t)$ of the whole system which has the
standard form
\begin{eqnarray}
\dot{\rho} &=&-i[H,\rho ]-\sum_{f=1}^{2}\frac{k_f}{2}\left(
b_{f}^{\dag }b_{f}\rho -2b_{f}\rho b_{f}^{\dag }+\rho b_{f}^{\dag
}b_{f}\right)
\nonumber \\
&&-\left[ \sum_{i=1}^{2}\frac{\kappa_i}{2}\left(a_{i,l}^{\dag
}a_{i,l}\rho -2a_{i,l}\rho a_{i,l}^{\dag }+\rho a_{i,l}^{\dag
}a_{i,l}\right) \right.
\nonumber \\
&&+\left.\sum_{i=2}^{3}\frac{\kappa_i}{2}\left(a_{i,r}^{\dag
}a_{i,r}\rho -2a_{i,r}\rho
a_{i,r}^{\dag }+\rho a_{i,r}^{\dag }a_{i,r}\right) \right]   \nonumber \\
&&-\sum_{i=1}^{3}\sum_{j=g_{0},g_{L},g_{R}}\frac{\gamma_{ij}
}{2}\left( S_{ij}^{\dag }S_{ij}^{-}\rho -2S_{ij}^{-}\rho
S_{ij}^{\dag }+\rho S_{ij}^{\dag }S_{ij}^{-}\right) ,
\end{eqnarray}
where $S_{ij}^{\dag }=|e\rangle _{i}\langle j|,$
$S_{ij}^{-}=|j\rangle _{i}\langle e|,$ $k_{f}$ $(\kappa_{i} )$
denotes the decay rate of the fibers (cavities) and $\gamma_{ij} $
is the spontaneous emission rate of the atoms. We assume $k_f=k$,
$\kappa_{i}=\kappa$ and $\gamma_{ij}=\gamma=\gamma_{0}/3$ for
simplicity. The fidelity $F$ versus the ratios $\Omega _{0}/g$ and
$\kappa /g$ $(\Omega _{0}/g$ and $\gamma /g)$ is displayed in the
top (bottom) panel of Fig. 5. We can see from Fig. 5 that with the
increasing of the laser intensity, the decoherence caused by the
atomic spontaneous emission is getting smaller and smaller, while
the decoherence caused by the cavity decay is becoming greater and
greater. The reason is that the adiabatic passage is just likely to
evolve within the dark state subspace under relatively large laser
intensity, and to violate the influence caused by the spontaneous
emission as well. However, the probabilities that the cavity fields
are excited increase with laser intensity, which in turn increase
the dissipation caused by the cavity decay finally. An appropriate
value $\Omega _{0}$ should be chosen by taking into account both the
factors (spontaneous emission of atoms and decay of cavities) as the
two error sources cannot be avoided simultaneously. The fidelity $F$
versus the cavity decay $\kappa /g$ and the fiber decay $k/g$ is
shown in Fig. 6, where we neglect the spontaneous emission of atoms.
As seen from the figure, the fidelity $F$ decreases with the
increasing cavity decay, but is almost unaffected by the fiber
decay. Even though we set $k/g=0.1$ (and $\gamma /g=\kappa /g=0)$
the fidelity is still as high as $F=0.998,$ so the decoherence due
to the fibers hardly influences the quality of the generated state.
The fidelity $F$ versus the decay of cavities $\kappa /g$ and the
spontaneous emission of atoms $\gamma /g$ is shown in Fig. 7 with
the fiber decay ignored. We see from Fig. 7 that $F$ decreases with
the increasing of both the cavity decay and the atomic spontaneous
emission. For a relative large value of $\gamma /g=\kappa /g=0.05,$
the fidelity is still about $F=0.850$ when the laser intensity
$\Omega _{0}/g=0.1$ chosen in our scheme. Therefore, our scheme is
robust in realistic conditions.

Finally, let us discuss on the experimental feasibility. The parameters $%
g=2\pi \times 75$ MHz, $\gamma =2\pi \times 2.62$ MHz and $\kappa =2\pi
\times 3.5$ MHz are achievable in optical cavities with the wavelength in
the region $630-850$ nm in recent experiments \cite{SMSTJKPRA05,JRBHJKPRA03}%
. A near-perfect fiber-cavity coupling with an efficiency larger than $%
97.20\% $ can be realized using fiber-taper coupling to high-Q
silica microspheres \cite{SMSTJKPRL03}. The optical fiber decay at a
$852$ nm wavelength is about $2.2$ dB/km
\cite{KJGVFIJQE04,SBZCPYPRA10}, which corresponds to the fiber decay
rate of $k=0.152$ MHz, which is lower than the cavity decay rate.
With these parameters, we will obtain a high fidelity $F,$ meaning
that it is possible to realize our scheme in a realistic experiment.

\section{Conclusion}

In conclusion, we have proposed a one-step scheme for deterministic
generation of GHZ states for any number of atoms individually trapped in
spatially separated cavities connected with short optical fibers via
adiabatic passage by appropriately tailoring the external driving laser
fields. The figure of merit is that we need not to precisely control the
generation time: the desired state emerges as a steady state of the system.
It is interesting that the generation time (i.e., the time it takes to
generate the GHZ state with a desired quality, i.e., $F=1-\varepsilon $ with
a predetermined small $\varepsilon )$ does not increase with the number of
atoms. This is of importance from the view point of decoherence when dealing
with a large-sized system such as entanglement of a big number of atoms. We
have also numerically calculated the influences of the driving laser's
parameters as well as of the decoherence effect caused by the atom
spontaneous emission and the cavity/fiber decay on the quality of the
generated state in terms of fidelity. The numerical results have revealed
that a relatively high fidelity of three-atom GHZ states can be obtained in
the presence of the above-mentioned influencing factors. Therefore, we do
hope that within the current experimental technology it would be possible to
realize our scheme.

\section{Acknowledgments}

S.Y.H and Y.X were supported by the National Natural Science Foundation of
China under grant no. 11047122 and no. 11105030, and China Postdoctoral
Science Foundation under grant no. 20100471450. N.B.A. was funded by Vietnam
National Foundation for Science and Technology Development (NAFOSTED) under
grant no. 103.99-2011.26.\newline

\end{document}